\documentclass[12pt,preprint]{emulateapj}

\def\GCN{{ GCN}}

\def\be{\begin{equation}}
\def\ee{\end{equation}}
\def\bea{\begin{eqnarray}}
\def\eea{\end{eqnarray}}

\usepackage{graphicx}

\begin{document}

\title{Is the late near-infrared bump in short-hard GRB 130603B due to
the Li-Paczynski kilonova?}

\author{Zhi-Ping Jin$^{1}$, Dong Xu$^{2}$, Yi-Zhong Fan$^{1}$, Xue-Feng Wu$^{3}$, and Da-Ming Wei$^{1}$}
\affil{$^1$ {Key Laboratory of Dark Matter and Space Astronomy, Purple Mountain Observatory, Chinese Academy of Science,
Nanjing, 210008, China}\\
$^{2}$ {Dark Cosmology Centre, Niels Bohr Institute, University of Copenhagen, Juliane Maries Vej 30, 2100 Copenhagen, Denmark}\\
$^{3}$ {Chinese Center for Antarctic Astronomy, Purple Mountain Observatory,
Chinese Academy of Sciences, Nanjing 210008, China}}
 \email{yzfan@pmo.ac.cn (YZF)}

\begin{abstract}
Short-hard gamma-ray bursts (GRBs) are widely believed to be
produced by the merger of two binary compact objects, specifically by
two neutron stars or by a neutron star orbiting a black hole.
According to the Li-Paczynski kilonova model, the merger would launch
sub-relativistic ejecta and a near-infrared/optical transient would
then occur, lasting up to days, which is powered by the radioactive
decay of heavy elements synthesized in the ejecta. The detection of a
late bump using the {\em Hubble Space Telescope} ({\em HST}) in the near-infrared
afterglow light curve of the short-hard GRB 130603B is indeed consistent
with such a model. However, as shown in this Letter, the limited {\em
HST} near-infrared lightcurve behavior can also be interpreted as the
synchrotron radiation of the external shock driven by a wide
mildly relativistic outflow. In such a scenario, the radio emission is
expected to peak with a flux of $\sim 100~\mu$Jy, which is detectable
for current radio arrays. Hence, the radio afterglow data can provide
complementary evidence on the nature of the bump in GRB 130603B. It is
worth noting that good spectroscopy during the bump phase in
short-hard bursts can test validity of either model above, analogous
to spectroscopy of broad-lined Type Ic supernova in long-soft GRBs.
\end{abstract}


\setlength{\parindent}{.25in}

\section{INTRODUCTION}
GRB 130603B triggered the Burst Alert Telescope (BAT) on board the {\it Swift} satellite at 15:49:14 UT on 2013 June 3
\citep{Melandri2013}. It had a $T_{90}$ duration of $0.18\pm 0.02$ s in the 15-350 keV band \citep{Barthelmy2013} and the BAT light curve reveals no trace of extended emission at the $\sim 0.005 ~{\rm counts~ det^{-1}~s^{-1}}$ level \citep{Norris2013}. The
spectral lag analysis reveals no significant delay of the high and low energy photons \citep{Norris2013}. All these facts together render GRB 130603B  a prototypical short-hard gamma-ray burst \citep[GRB;][]{deUgartePostigo2013, Bromberg2013}. GRB 130603B is the first short GRB with absorption spectroscopy \citep{deUgartePostigo2013}.  The other remarkable discovery made in GRB 130603B is an infrared bump appearing at $t\sim 9$ days after the burst \citep{Tanvir2013,Berger2013}, which has been interpreted as the infrared/optical transient powered by the radioactive decay of heavy elements synthesized in the sub-relativistic ejecta launched by either the neutron star binary merger or the neutron star-black hole merger, i.e., the Li-Paczynski kilonova (e.g., Li \& Paczynski 1998; Rosswog 2005; Metzger et al. 2010; Goriely et al. 2011; Kasen et al. 2013; Barnes \& Kasen 2013; Tanaka \& Hotokezaka 2013; Grossman et al. 2013). If the  Li-Paczynski kilonova origin is established, the neutron star binary merger or the neutron star-black hole merger origin of some (if not all) short GRBs, a model proposed in 1990s (Eichler et al. 1989; Narayan et al. 1992), will be confirmed. Hence, some short and long GRBs do have very different physical origin, as widely speculated (Narayan et al. 2001). In view of the fundamental importance of such a kind of interpretation, it is necessary to check whether other possibilities exist, and if they do, how these possibilities can be further constrained. That is the main purpose of this letter.

\section{The difficulty of interpreting the late infrared bump as the regular afterglow}
Tanvir et al. (2013) suggested that there are two reasons against the regular afterglow origin of the late infrared bump of GRB 130603B. One is that the optical afterglow lightcurve of GRB 130603B drops with time more quickly than $t^{-2}$ for $t>10$ hours after the trigger of the burst. The near-infrared flux, on the other hand, is in excess of the same extrapolated power
law (see Figure 2 of Tanvir et al. 2013). The other is the significant color evolution
of the transient, defined as the difference between the magnitudes
in each filter, which evolves from $R_{606}-H_{160}\approx 1.7\pm 0.15$ mag at about
14 hr to greater than $R_{606}-H_{160}\approx 2.5$ mag at $\sim 9$ days. Below we discuss the afterglow model extensively and demonstrate that in the regular fireball afterglow model no color evolution should be present in the time interval of $~0.6-9$ days, which is in support of Tanvir et al. (2013)'s argument.

As found in \citet{deUgartePostigo2013}, assuming a spectral break $\nu^{-0.5}$ between the optical and X-ray bands as that expected in the standard afterglow model (Piran 1999), the near-infrared to X-ray spectral energy distribution (SED) of GRB 130603B 8.5 hr after the burst onset can be nicely fitted by an extinction of $A_{\rm V}=0.86$ mag and a Small Magellanic Cloud (SMC) extinction law. The SED is well fitted with a spectral break at $\nu_{\rm break}\approx 10^{16}~{\rm Hz}$ and the lower and high energy spectral indexes are $\alpha_{\rm O}=-0.65\pm 0.09$ and $\alpha_{\rm X}=-1.15\pm 0.11$, respectively. Using the optical afterglow data at $t\sim 0.6$ day (Tanvir et al. 2013) and the public X-ray afterglow data, we obtain a very similar SED but $\nu_{\rm break}$ shifts to $\sim 6\times 10^{15}$ Hz (see Figure 1). The optical and X-ray spectra suggest that the break frequency is the so-called cooling frequency $\nu_{\rm c}$ in the fireball afterglow model (Piran 1999). For the burst born in stellar wind, $\nu_{\rm c} \propto t^{1/2}$, i.e., the later the observation, the higher the cooling frequency. The SEDs at $t\sim 0.35$ day and $0.6$ day are not in support of such a tendency. Instead they are in agreement with the case of that $\nu_{\rm c} \propto t^{-1/2}$ for the burst born in the ISM-like medium. Hence, $\nu_{\rm c} \sim 10^{16}(t/0.35~{\rm day})^{-1/2} \sim 2\times 10^{15}$ Hz at $t\sim 9$ days, which is still well above the optical band. Such results holds as long as the ejecta sideways expansion is unimportant. If the sideways expansion is important, in both wind- and ISM-like medium models, $\nu_{\rm c} \propto t^{0}$. All these facts together rule out the presence of a significant color evolution of the infrared/optical afterglow emission in the time interval of $0.6-9$ day. Therefore the observed soft infrared/optical emission at $t\sim 9$ days should have a different physical origin. Tanvir et al. (2013) and Berger et al. (2013) have interpreted the infrared bump as the Li-Paczynski kilonova. This kind of interpretation is extremely attractive. However, the {\em Hubble Space Telescope} ({\em HST}) data is very rare and other scenarios should also been investigated.

\begin{figure}
\includegraphics[angle=0,width=0.5\textwidth]{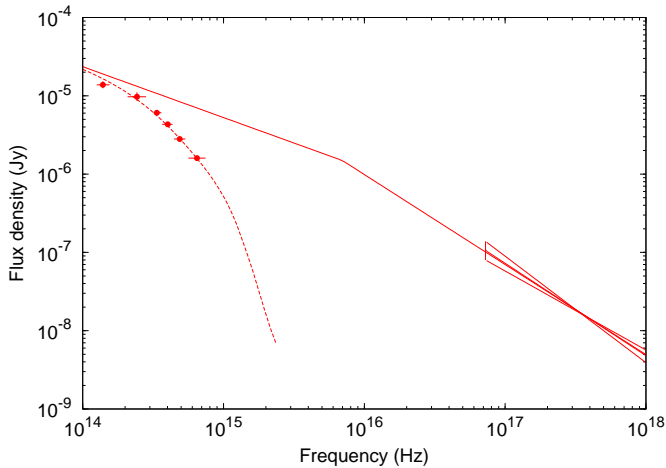}
\caption{SED fit to the afterglow of GRB 130603B. The red solid line is the 0.6 day intrinsic broken power-law spectra with indexes 0.65 and 1.15 and the break frequency 6.0$\times 10^{15}$Hz. The dashed line is the extinct spectrum with $A_{\rm v}$=0.9 for the host galaxy (with SMC extinction law) and the Galactic $A_{\rm v}$=0.06. The optical data and 3$\sigma$ upper limits are taken from Tanvir et al. (2013) and the XRT data is from http://www.swift.ac.uk/xrt$_{-}$curves/00557310/.} \label{fig:SED}
\end{figure}

\section{The second-component jet model for the infrared bump of GRB 130603B?}
Two component jet model has been adopted to interpret some peculiar afterglow emission of both long and short GRBs (for the former, see, e.g., Berger et al. 2003; Huang et al. 2004; Racusin et al. 2008; for the latter, see Jin et al. 2007). In such a model, the narrow energetic core produce prompt $\gamma-$ray emission and then the early bright afterglow emission while the much wider but less energetic ejecta component will emerge at a late time, depending on its bulk Lorentz factor. The infrared bump likely peaks at $t\sim 9$ days after the burst onset, then the initial Lorentz factor of the ejecta should satisfy
\begin{equation}
\Gamma_0 \sim 2~E_{\rm k,w,50}^{1/8}(t/9~{\rm day})^{-3/8}n_0^{-1/8},
\end{equation}
where $E_{\rm k,w}$ is the kinetic energy of the mildly relativistic outflow component and $n$ is the number density of the circum-burst medium (for simplicity, below we just discuss the ISM-like medium that is favored by the SEDs). Note that here and throughout the text the convenience $Q_{\rm x}=Q/10^{\rm x}$ has been adopted except for specific notations.

We also point out that a mildly relativistic outflow component is {\it not} unexpected. For example, in both the double neutron star merger scenario the neutron star-black hole merger scenario, a wide but mildly relativistic outflow surrounding the ultra-relativistic GRB ejecta may be formed as a result of the interaction of the outflow with the surrounding material (e.g., Aloy et al. 2005). After the merger of the double neutron stars, a supramassive/stable magnetar rather than a black hole may be formed (e.g., Gao \& Fan 2006; Zhang 2013; Giacomazzo \& Perna 2013). The wind of the magnetar that possibly suffers from significant kinetic energy loss via gravitational wave radiation (Fan et al. 2013) may be able to accelerate the material ejected from the double neutron star merger to a mildly relativistic velocity as well (Fan \& Xu 2006; Gao et al. 2013).

The cooling Lorentz factor of the external forward shock electrons can be estimated as $\nu_{\rm c} \approx 10^{16}~{\rm Hz}~E_{\rm k,51}^{-1/2}\epsilon_{\rm B,-2}^{-3/2}n_0^{-1}(t/1~{\rm day})^{-1/2}(1+z)^{-1/2}$ (Piran 1999), where $\epsilon_{\rm B}$ is the fraction of shock energy given to the magnetic field and $z=0.356$ is the redshift of GRB 130603B \citep{deUgartePostigo2013}. For the narrow and wider ejecta components, the number density of the medium should be the same and the initial kinetic energy is expected to be different and usually we have $E_{\rm k,n}>E_{\rm k,w}$. As mentioned above, for the narrow ejecta component $\nu_{\rm c,n}\sim 10^{16}$ Hz at $t\sim 0.35$ day, hence
\begin{equation}
\nu_{\rm c,w}\sim 10^{16}~{\rm Hz}~({E_{\rm k,n}\over E_{\rm k,w}})^{1/2}({\epsilon_{\rm B,n}\over \epsilon_{\rm B,w}})^{3/2}({t\over 0.35~{\rm day}})^{-1/2}.
\end{equation}
To interpret the identified softness of near-infrared bump (i.e., $\Delta (R_{606}-H_{160})\approx 0.8\pm 0.15$ mag), the synchrotron radiation spectrum of the second-component ejecta should be softer than that of the early ($t\sim 0.6$ day) afterglow by a factor of $\nu^{-0.75\pm 0.14}$. The required power-law distribution index of the electrons accelerated by the wide-component ejecta is $p_{\rm w} \sim 2.8\pm 0.3$ as long as $\nu_{\rm c,w}<\nu_{_{\rm F{606}W}}$ at $t\geq 9$ days (where $p_{\rm n}\sim 2.3$ has been adopted), for which we need
\begin{equation}
\epsilon_{\rm B,w}\geq 5.4\epsilon_{\rm B,n}~({E_{\rm k,n}\over 10~E_{\rm k,w}})^{1/3}.
\end{equation}
It is unclear why the narrow and wide outflow components have different $\epsilon_{\rm B}$ (possibly also $\epsilon_{\rm e}$ and/or $p$). However, we note that the best-fitted microphysical parameters of GRBs differ from burst to
burst (Panaitescu \& Kumar 2001) and no universal values have been obtained. Moreover, in the modeling of the afterglow emission of some GRBs within the two-component jet scenario, the best-fitted microphysical parameters are found to be different for the narrow and wide components (e.g., Jin et al. 2007; Racusin et al. 2008). Hence, we suggest that the request $\epsilon_{\rm B,w}> \epsilon_{\rm B,n}$ is reasonable and possible.

Simultaneous with the $\sim 0.2~{\rm \mu Jy}$ infrared emission, the radio radiation flux is expected to be
\begin{eqnarray}
F_{\nu_{\rm radio}} \geq  0.2~{\rm \mu Jy}~({\nu_{\rm radio}\over 1.8\times 10^{14}~{\rm Hz}})^{-(p_{\rm w}-1)/2}\sim 340 {\rm \mu Jy},
\end{eqnarray}
for $p_{\rm w}\sim 2.5$ and $\nu_{\rm radio}\sim 10^{10}~{\rm Hz}$. This is because both the typical synchrotron radiation frequency $\nu_{\rm m,w}\approx 4\times 10^{9}~{\rm Hz}~E_{\rm k,w,50}^{1/2}\epsilon_{\rm B,w,-1}^{1/2}\epsilon_{\rm e,w,-1}^{2}(t/9~{\rm day})^{-3/2}$ and the synchrotron self-absorption frequency $\nu_{\rm a,w}\approx 5.7\times 10^{9}~{\rm Hz}~E_{\rm k,w,50}^{0.35}\epsilon_{\rm B,w,-1}^{0.35}n_0^{0.3}\epsilon_{\rm e,w,-1}^{0.46}(t/9~{\rm day})^{-0.73}$ are below $\sim 10^{10}~{\rm Hz}$.
Note that even for $p_{\rm w} \sim 2.3$, we have $F_{\nu_{\rm radio}}\sim 100~\mu$Jy. Such a flux is bright enough to be reliably detected by some radio arrays (for example, the Karl G. Jansky Very Large Array) in performance. The non-detection will in turn impose a tight constraint on the external forward shock radiation origin of the infrared bump.

To better show our idea, we calculate the flux numerically. The code used here has been developed in Fan \& Piran (2006) and Zhang et al.
(2006). The dynamical evolution of the outflow is calculated
using the formulae in Huang et al. (2000), which can be
used to describe the dynamical evolution of the outflow for both
the relativistic and non-relativistic phases. The energy distribution of the shock-accelerated electrons is calculated
by solving the continuity equation with the power-law source
function $Q=K\gamma_{\rm e}^{-p_{\rm w}}$, normalized by a local injection rate (Moderski et al. 2000). The cooling
of the electrons due to both synchrotron and inverse Compton
 has been taken into account.

In Figure 2, we have presented one numerical example which can fit the limited {\em HST} data of GRB 130603B. The physical parameters adopted in the fit are as follows: $\epsilon_{\rm e,w}=0.1$, $\epsilon_{\rm B,w}=0.1$, $p_{\rm w}=2.5$, $n=1.0~{\rm cm}^{-3}$, $E_{\rm k,w}=4\times 10^{50}~{\rm erg}$, the initial Lorentz factor of the outflow $\Gamma_0=3.0$, and the half-opening angle is assumed to be $\theta_j=1.0$. As one can see both the temporal and spectral properties of the infrared bump of GRB 130603B can be reproduced. The radio afterglow emission is so bright ($\sim 200\mu$Jy) that can be well detected by Karl G. Jansky Very Large Array-like telescopes. The non-detection would impose a tight constraint on the mildly relativistic outflow model.

\begin{figure}
 \includegraphics[angle=0,width=0.5\textwidth]{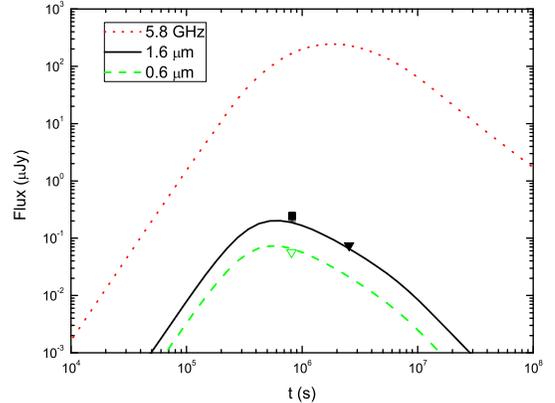}
\caption{Multi-wavelength afterglow emission of a wide mildly relativistic outflow component, which is consistent with the {\em HST} data (extinction corrected) of the near-infrared bump of GRB 130603B (Tanvir et al. 2013).}
\label{fig:fit}
\end{figure}

\section{Discussion}
Our understanding of short GRBs, a kind of $\gamma-$ray outbursts with a duration less than two seconds \citep{Kouveliotou1993}, has been revolutionized due to the successful performance of the {\it Swift} satellite. For a good fraction of short GRBs, the binary-neutron-star or the black hole-neutron star merger model \citep{Eichler1989,Narayan1992} has been supported by their
host galaxy properties and by the non-association with
bright supernovae. A smoking-gun signature, i.e., a supernova-like infrared/optical transient powered by the radioactive decay of heavy elements synthesized in the ejecta launched by either the neutron star binary merger or the neutron star-black hole merger (i.e., the Li-Paczynski kilonova), however, has not yet been unambiguously identified. The best candidate of such a smoking-gun signature is likely the infrared bump detected at $t\sim 9$ days after the onset of GRB 130603B \citep{Tanvir2013,Berger2013}. If the  Li-Paczynski kilonova origin is confirmed, the neutron star binary merger or alternatively the neutron star-black hole merger origin of some (if not all) short GRBs will be established. Hence, some short and long GRBs do have very different physical origin \citep{Narayan2001}. In view of the fundamental importance, it is necessary to check whether or not other possible physical origins of the infrared bump of GRB 130603B exist. As shown in this Letter the {\em HST} near-infrared data is very limited and can be interpreted as the synchrotron radiation of the external shock driven by a wide mildly relativistic outflow. Interestingly, a wide mildly relativistic outflow associated with the ultra-relativistic GRB ejecta has been ``observed" in the numerical simulation (e.g., Aloy et al. 2005) or ``predicted" in the magnetar central engine model of short GRBs (e.g., Fan \& Xu 2006; Gao et al. 2013). In the mildly relativistic outflow model, the radio emission is expected to peak at $t\sim 10^{6}$ s (see Fig.2) with a flux $\sim 100~\mu$Jy, which is detectable for some radio arrays in performance. While the synchrotron radio radiation powered by the kilonova outflow is expected to peak at $t\sim 8\times 10^{8}(V_{\rm nova}/0.1~c)^{-5/3}E_{\rm nova,51}^{1/3}n_0^{-1/3}$ s, where $E_{\rm nova}$ ($V_{\rm nova}$) is the kinetic energy (velocity) of the kilonova outflow and $c$ is the speed of light. Hence, the radio afterglow data can provide
complementary evidence on the nature of the near-infrared bump in GRB 130603B or similar events in the future. It is
worth noting that good spectroscopy during the bump phase in
short-hard bursts can test validity of either model above, analogous
to spectroscopy of broad-lined Type Ic supernova in long-soft GRBs.

If the mildly relativistic outflow model has been confirmed by (future) observations, the near-infrared bump like that detected in GRB 130603B is still valuable for those interested in searching for the electromagnetic counterparts of the merger of two neutron stars or a neutron star and a black hole since such a kind of signal is expected to be almost isotropic due to its low bulk Lorentz factor. As shown in the numerical simulation (Aloy et al. 2005), the mildly relativistic outflow may be common, thus the observational prospect can not be ignored.

\section*{Acknowledgments}
This work was supported in part by the 973 Programme of China under grants 2009CB824800 and 2013CB837000, the National Natural Science of China under grants 11073057, 11103084 and 11273063, and the Foundation for Distinguished Young Scholars of Jiangsu Province, China (No. BK2012047).  Y.Z.F. and X.F.W. are also supported by the 100 Talents programme of Chinese Academy of Sciences. D.X. acknowledges support from the ERC-StG grant EGGS-278202 and IDA.

\clearpage

\end{document}